\documentclass[final,1p,times]{elsarticle} 
\usepackage{graphicx} 
\usepackage{amssymb} 
\usepackage{amsthm} 
\usepackage{lineno} 
\newcommand{\lsim}{\,{\buildrel < \over {_\sim}}\,}
\newcommand{\gsim}{\,{\buildrel > \over {_\sim}}\,}

\journal{Nuclear Physics A} 
\begin{document} 

\begin{frontmatter} 


\title{Measurements of jet structure and fragmentation from full jet reconstruction in heavy ion collisions at RHIC}

\author{Elena Bruna for the STAR Collaboration}

\address{Physics Department, Yale University, New Haven, CT 06520, U.S.A}

\begin{abstract} 
Measurements of inclusive hadron suppression and di-hadron azimuthal
correlations have provided
important insights into jet quenching in hot QCD matter. However, they do not provide access to the energy of the hard scattering and are limited in their sensitivity since they can be affected by biases toward hard fragmentation and small energy loss. Full jet
reconstruction in heavy-ion collisions enables a complete study of the modification of jet
structure due to energy loss, but is challenging due to the high-multiplicity environment.
We present measurements of fully reconstructed di-jets at mid-rapidity
in 200 GeV p+p and central Au+Au collisions performed with the STAR detector.
We compare fragmentation functions measured in 200
GeV p+p and central Au+Au collisions and assess the systematic
uncertainties of their ratio. 
\end{abstract} 

\end{frontmatter} 


\section{Introduction}
There has been recent significant progress in high-p$_T$ physics at RHIC with the success of full jet reconstruction in heavy-ion collisions~\cite{joern,sevil}.
Unbiased reconstruction of jets provides experimental access to the initial hard scattering, independent of the presence of the nuclear medium.
Therefore, a binary scaling of jet production from p+p to Au+Au is expected.
Insight into the jet's structure is provided by studying fragmentation functions (FF).
In-medium softening of the FF with respect to p+p reference measurements, i.e. a distortion of the so called ``hump-backed'' plateau~\cite{wiedemann,borghini}, should be observable in Au+Au with an unbiased jet population.
The large background and its fluctuations make full jet reconstruction a challenge in the high-multiplicity environment at RHIC. We performed jet measurements in STAR with the modern jet-finding techniques and utilizing data-driven correction schemes. 

\section{Experimental techniques and analysis}
The STAR sub-detectors used for jet reconstruction are the Time Projection Chamber (TPC) for charged particles and the Barrel Electromagnetic Calorimeter (BEMC) for the neutral energy. Both TPC and BEMC have full azimuthal coverage and pseudo-rapidity acceptance $|\eta|<1$.
Corrections for double-counting of electrons and hadronic energy deposition in the BEMC are applied. 
The analysis presented in these proceedings is based on p+p year 2006 and 0-20$\%$ most central Au+Au year 2007 events. Both data sets were selected with an online High-Tower (HT) trigger in the BEMC which requires the transverse energy in a tower to be  $E_T>5.4$ GeV.

The goal of jet-finding algorithms is to cluster charged tracks and neutral towers into jets in the $\eta-\phi$ plane.
The analysis reported in these proceedings utilizes the ``anti-$k_t$'' algorithm~\cite{antikt}, which is a recombination algorithm and part of the FastJet package~\cite{fastjet}. 
 Like all recombination algorithms, ``anti-$k_t$'' is collinear and infrared safe. ``Anti-$k_t$'' is used in this analysis because it is expected to be less sensitive to background effects in heavy-ion collisions.

The di-jet analysis presented here is performed on events with a ``trigger'' jet, which matches the online triggered tower in the BEMC, and a ``recoil'' jet on the away side of the trigger jet (i.e. $\Delta \phi \sim \pi$).
A critical issue in jet analyses in Au+Au is the background.
FastJet provides an estimate of the background $p_t$ per unit area, that is subtracted to get the jet component p$_{t,rec}^{jet}$.
The background energy is of the order of 45 GeV in a cone radius of R=0.4 ($R=\sqrt{\Delta \phi^2 + \Delta \eta^2}$). 
The background exhibits significant fluctuations in a central Au+Au event. We parameterize the upwards fluctuations of the background by a Gaussian with width $\sigma$ of the order of 6-7 GeV~\cite{joern,mateusz}.
A resolution parameter $R=0.4$ was chosen, to suppress the background without losing a large fraction of the jet energy ($\sim$80$\%$ of the jet energy lies within R=0.4 for 20 GeV p+p jets~\cite{helen}).
The background fluctuations can be suppressed by requiring a minimum transverse momentum ($p_t^{cut}$) for a particle to be included in the jet. 
For the trigger jet a $p_t^{cut}=2$ GeV/c was applied in order to have the same energy scale in p+p and Au+Au. A $p_t^{cut}$ introduces a strong bias in the jet population, therefore the recoil jets are reconstructed with a minimal  $p_t^{cut}=0.15$ GeV/c due to the TPC acceptance. The requirement of $p_{t,rec}^{jet}(trigger)>10$ GeV/c for the reconstructed trigger jet minimizes the contribution of ``fake'' jets. 
Other sources of background in di-jet analyses are (a) ``fake'' jets and (b) hard scatterings uncorrelated to the di-jet.
Additional hard scatterings uncorrelated to the one that produced the di-jet pair frequently occur in heavy-ion collisions.
The background di-jet coincidence rate is estimated from the spectrum of associated jets located at $\sim \pi/2$ with respect to the trigger jet. This spectrum is used to correct for background jets.

A similar analysis on fully reconstructed jets performed on untriggered p+p and Au+Au events is reported in~\cite{mateusz}.

\section{Di-jet coincidence rate and recoil fragmentation functions}

The data shown in these proceedings were not corrected for the trigger efficiency and for the difference in tracking efficiency between p+p and Au+Au. An absolute correction to get the parton energy will be performed in the future.
A data-driven unfolding of the background fluctuations from the Au+Au di-jet spectra and FF of recoil jets was applied to allow a direct comparison of p+p and Au+Au. 
The unfolding procedure removes the artificial hardening of the spectra due to the convolution of the background fluctuations with the steeply falling jet spectrum.

\begin{figure}
\centering

\resizebox{0.48\textwidth}{!}{  \includegraphics{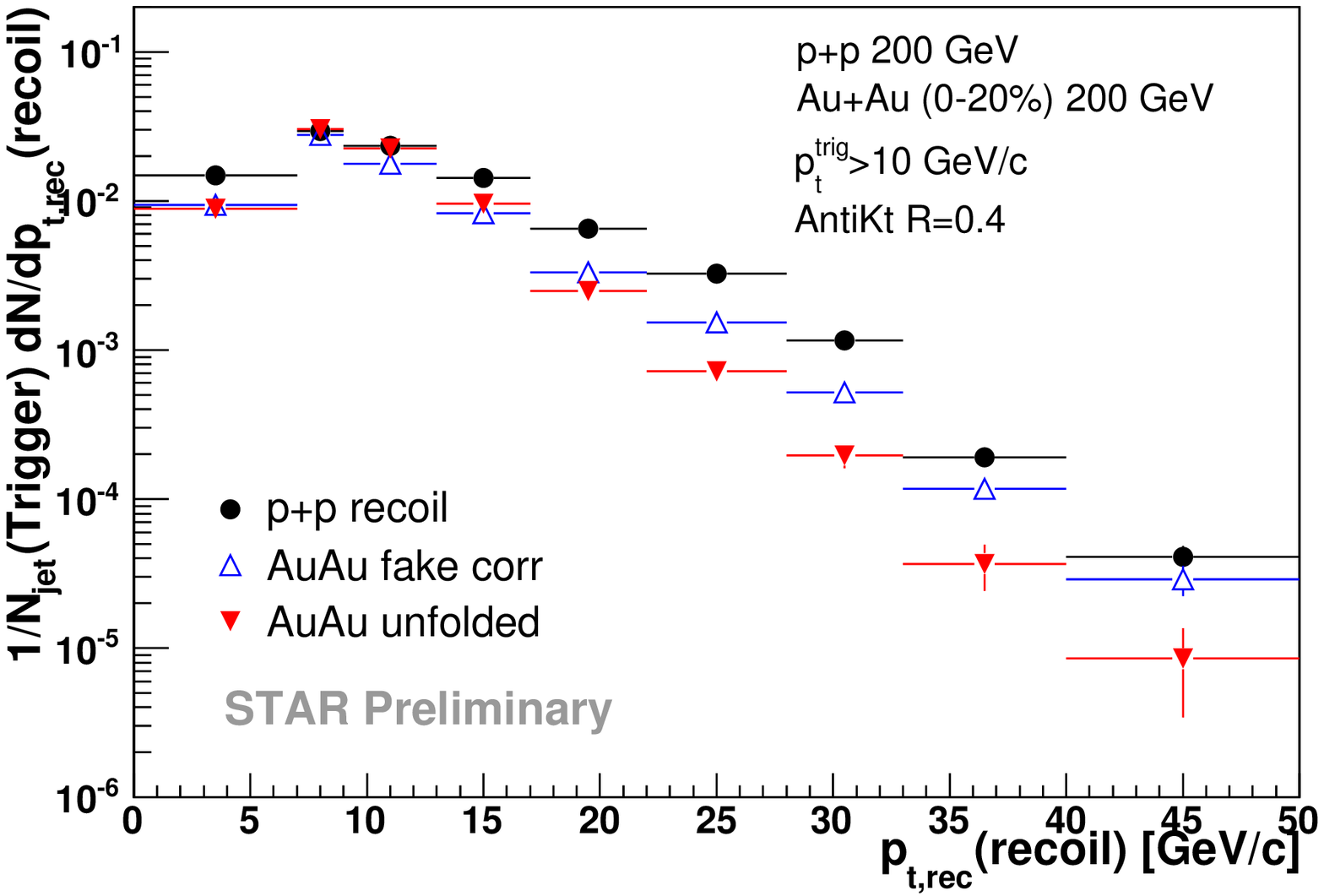}}
\resizebox{0.48\textwidth}{!}{  \includegraphics{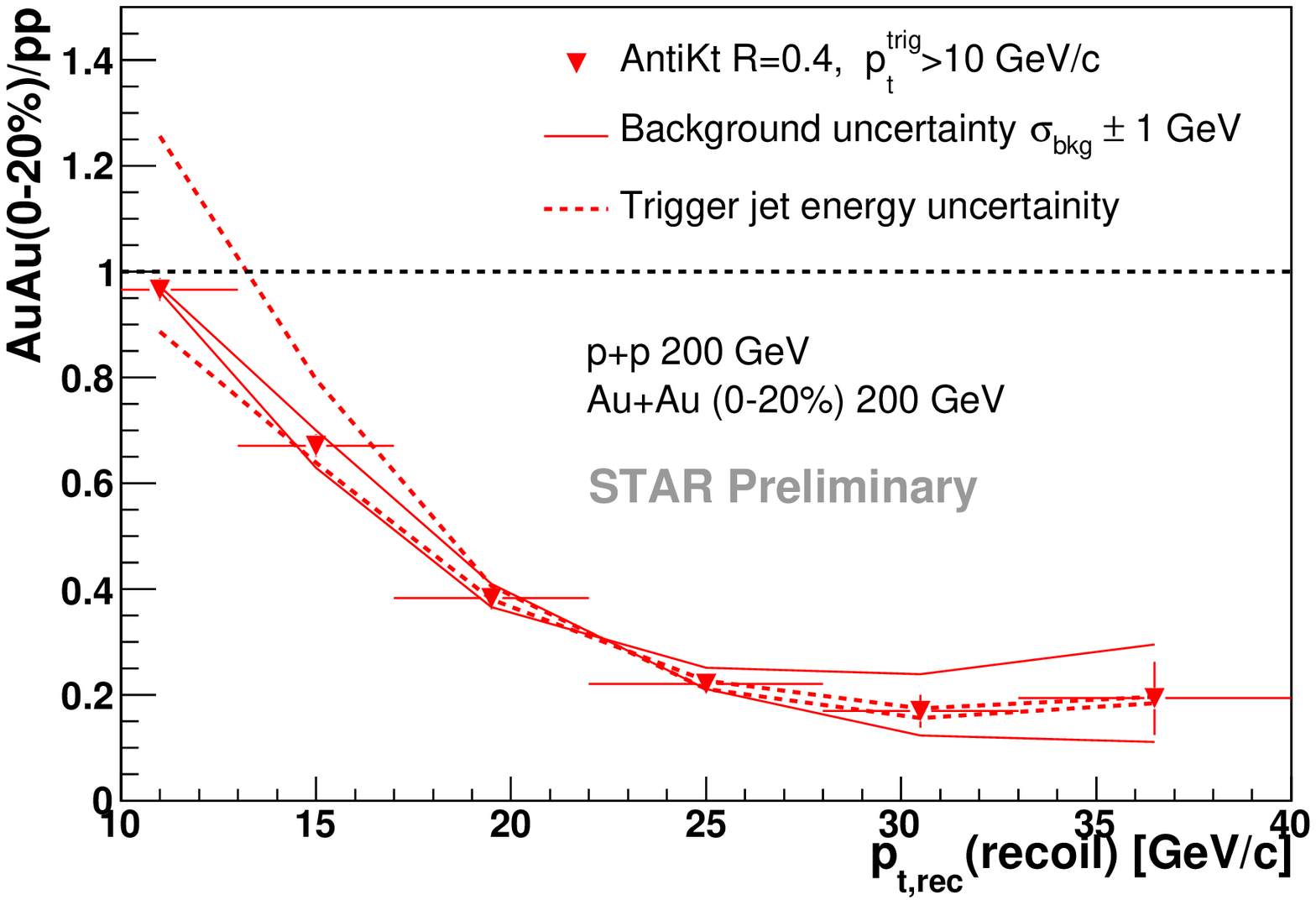}}

\caption{Left plot: $p_t$ spectra of recoil jets (for $p_t^{trig}>10$ GeV/c) in p+p (close circles) and in Au+Au after the correction for fake jets (open triangles) and unfolding of background fluctuations (close triangles). The spectra are normalized to the number of trigger jets. No corrections for trigger and tracking efficiency were applied. Right plot: ratio of $p_t$ spectra of recoil jets in Au+Au (corrected for fake jets and background fluctuations) to p+p. The curves indicate the systematic uncertainties in the estimation of the background fluctuations (solid) and of the $p_{t}$ of the trigger jet assuming a jet-$p_t$ resolution of 25$\%$ (dotted)~\cite{helen}.}
\label{fig:ppAuAuDiJets}
\end{figure}

The recoil jet spectra in p+p and Au+Au normalized to the number of trigger jets are shown in Fig.~\ref{fig:ppAuAuDiJets} (left). 
The ratio of the di-jet spectra of Au+Au to p+p is reported in Fig.~\ref{fig:ppAuAuDiJets} (right), indicating a strong suppression of recoil jets in Au+Au with respect to p+p at a given reconstructed jet energy.
This is in constrast with the expected value of unity for unbiased jet reconstruction. 

We also report measurements of the fragmentation functions for recoil jets and compare the results in p+p and Au+Au. In contrast to trigger jets, recoil jets are not affected by the trigger bias that artificially enhances the neutral energy component in the jet firing the trigger.
In order to include the soft jet fragments, the FF are measured from charged hadrons in a radius of 0.7 around the jet axis, while the reconstructed jet energy is from R=0.4.
The background component of the FF in Au+Au is estimated on an event-by-event basis from the charged particle spectrum in the area outside the two jets with the highest reconstructed energy.





\begin{figure}
\centering

\resizebox{0.48\textwidth}{!}{  \includegraphics{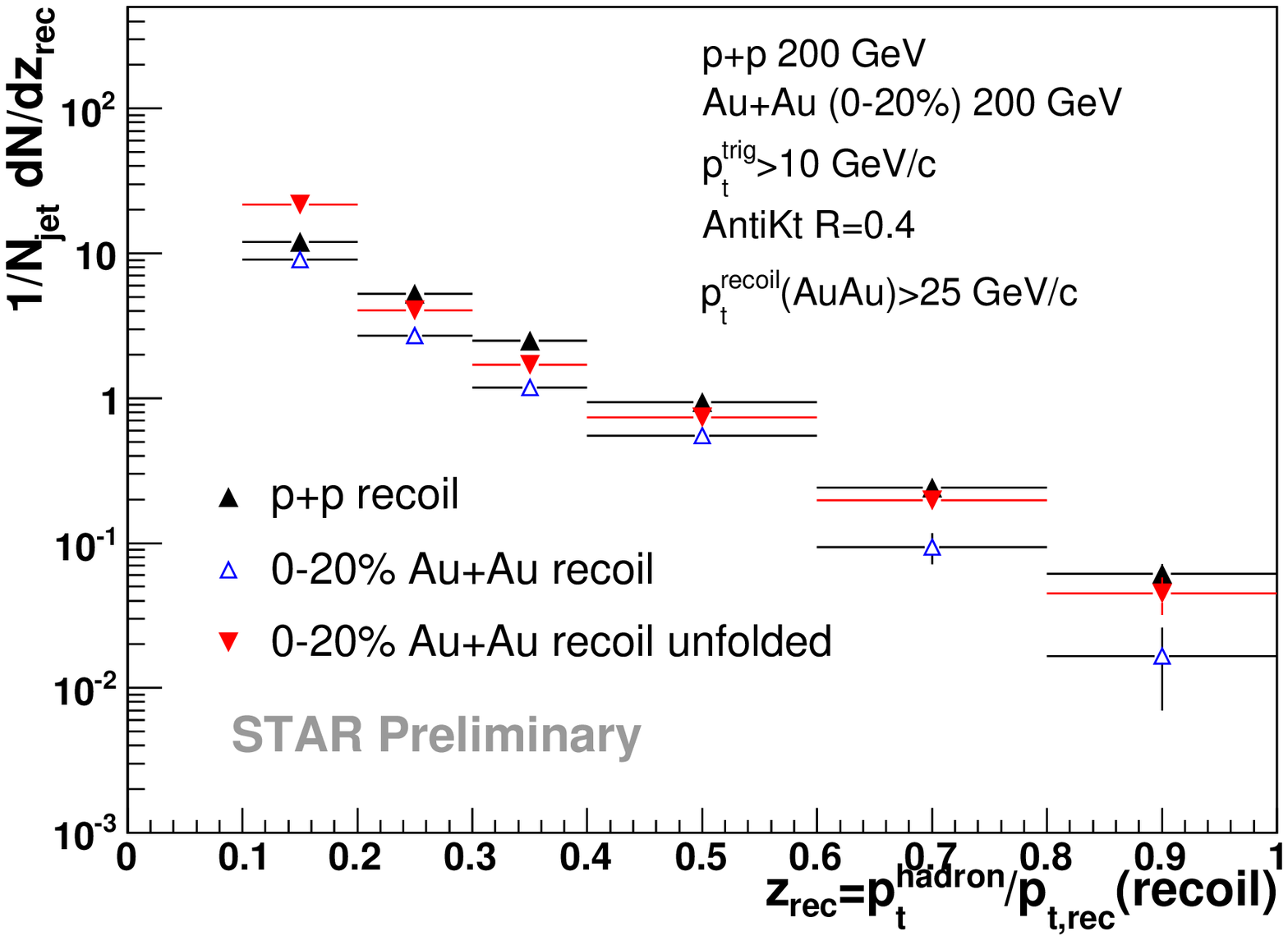}}
\resizebox{0.48\textwidth}{!}{  \includegraphics{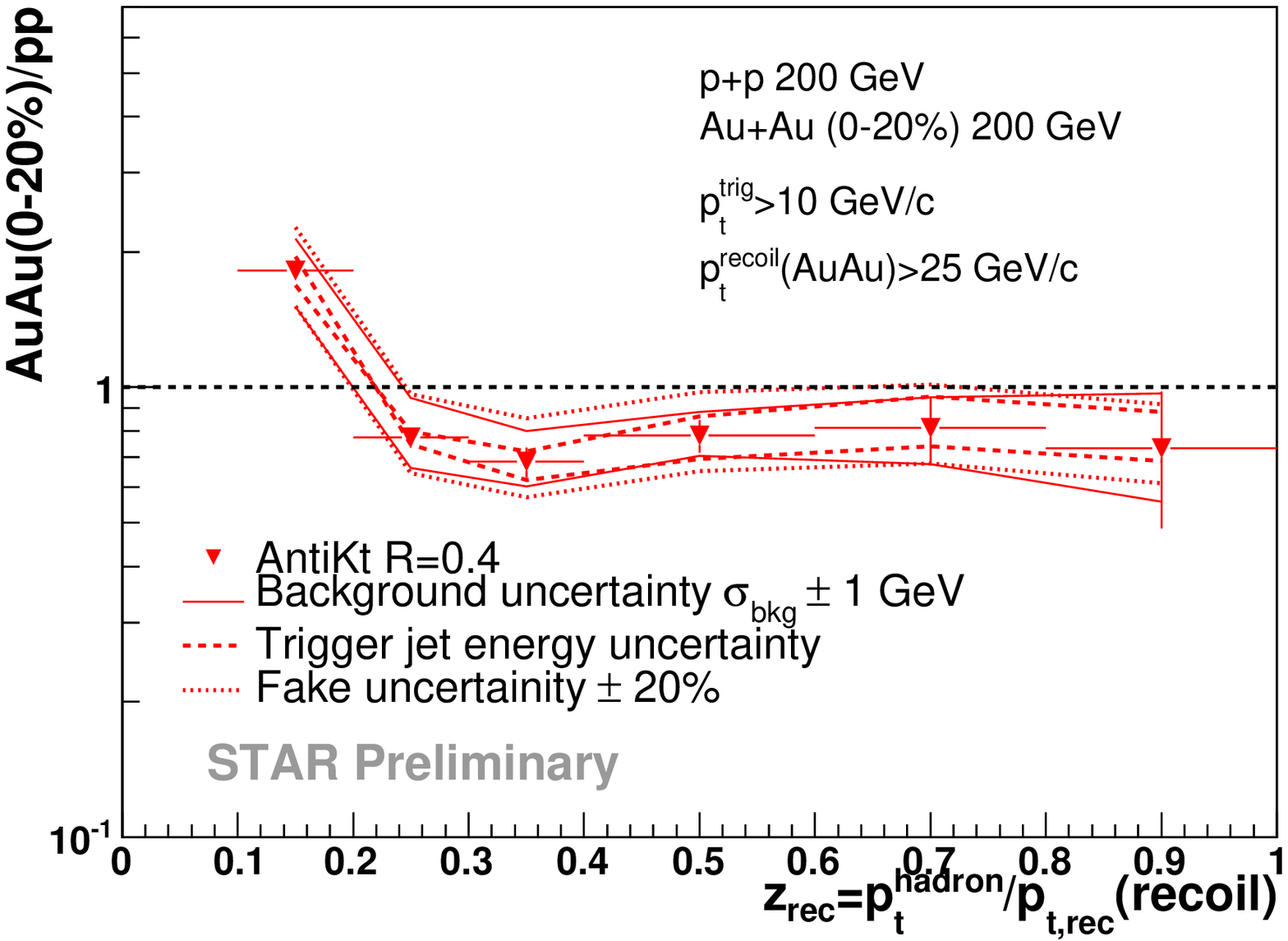}}

\caption{Left plot: FF for recoil jets (normalized to the number of recoil jets) in p+p, Au+Au (open triangles), Au+Au after unfolding for $p_{t,rec}^{recoil}(AuAu)>25$ GeV/c and $p_{t,rec}^{trig}>10$ GeV/c.  Right plot: ratio of the $z$ distributions measured in 0-20$\%$ central Au+Au events to p+p collisions for $p_{t,rec}^{recoil}(AuAu)>25$ GeV/c. The curves indicate the systematic uncertainties in the estimation of the background fluctuations (solid) and of the trigger jet energy (dotted). The uncertainty due to the estimation of fake jets (small dotted) was assumed to only affect the normalization of the $z$ distributions.}
\label{fig:ppAuAuFF}
\end{figure}

The FF of reconstructed recoil jets are shown for $p_{t,rec}^{recoil}(AuAu)>25$ GeV/c (Fig.~\ref{fig:ppAuAuFF}). The p+p jets corresponding to  $p_{t,rec}^{recoil}(AuAu)>25$ GeV/c are selected taking the background fluctuations into account. The mean jet $p_t$ of the selected p+p jets is $p_{t,rec}^{recoil}(pp)\simeq 25$ GeV/c. 
The results of the unfolding procedure applied to the $z$ distributions ($z=p_t^{hadron}/p_{t,rec}^{recoil}$) are shown in Fig.~\ref{fig:ppAuAuFF} (left).
The FF are harder since the artificial hard contribution in the jet spectrum due to the background fluctuations has been removed after the unfolding.
The ratio of the $z$ distributions of Au+Au to p+p is shown in Fig.~\ref{fig:ppAuAuFF} (right). No significant modification of the fragmentation functions of recoil jets is observed with $p_{t,rec}^{recoil}(AuAu)>25$ GeV/c for $z\gsim 0.2$, in contrast to the expectation of a softening of the FF for an unbiased jet population and compared to the measured suppression of high-p$_t$ hadrons at RHIC ($R_{AA} \sim 0.2$~\cite{Raa}). The low-$z$ part of the fragmentation function ($z\lsim 0.1$) is still under investigation since it is dominated by the underlying background in Au+Au and therefore might be affected by a large uncertainty due to background subtraction.

\section{Discussion of the results}

 Due to the strong trigger bias in the events used in this analysis, a particular class of di-jets was selected. Indeed, the trigger jets seem to be unmodified and therefore most likely come from the surface of the medium, which would bias the recoil jets towards a maximum in-medium path length and to a maximum energy loss effect.

Given this ``extreme'' selection of recoil jets, we reported a significant suppression of di-jet coincidence rates in central Au+Au with respect to p+p at a given reconstructed jet energy. In addition, we observed the absence of strong modification of the FF of recoil jets in 0-20$\%$ Au+Au with $p_{t,rec}^{recoil}(AuAu)>25$ GeV/c with respect to p+p.


The above observations can be explained via a scenario where the jet is broadened and its energy is not fully recovered in $R=0.4$ with respect to p+p with the current jet-finding algorithms.
In this case a shift of the recoil spectrum towards smaller energies would be expected (and would explain the results in Fig.~\ref{fig:ppAuAuDiJets}). 
A softening of the FF is expected in Au+Au to account for the measured high-p$_t$ hadron suppression.
In a  jet broadening scenario, the observed absence of a strong modification of the measured FF in Au+Au (Fig.~\ref{fig:ppAuAuFF}, right) could be due to an underestimation of the energy in R=0.4. 
To prove this scenario we must confirm whether or not jet broadening occurs beyond R=0.4. If this is the case, a measure of the residual out-of-cone jet energy would then allow full jet energy reconstruction and a measurement of how the FF is modified.

In the case that a part of the jet population is broadened to the extent that it cannot be recovered, we would not expect a modification in the jet shape for the measured recoil jets. The measured FF in Au+Au would be those of the surviving recoil jets, i.e. the ones that only minimally interact without being absorbed, such as the ones emitted tangentially to the surface of the medium.
To test this scenario, the di-jet coincidence rates for different radii will be measured and their ratio should be the same in p+p and Au+Au.

The corrections and systematic error bands reported do not yet extensively assess the systematic uncertainties in the di-jet analysis. Furthermore, detailed systematic studies (i.e. tracking efficiency, calorimeter calibration, missing neutral energy, etc.) are being conducted in order to access the kinematics of the hard scattering.
The jet-energy profile will also be investigated in order to improve our understanding of the mechanism of jet broadening in the medium.




\end{document}